\documentstyle[11pt,fleqn]{article}
\def\ref{par\noindent\hangindent=6mm\hangafter=1}
\begin{document}
\vbox{
\rightline{IFUG-95/08 r}
\rightline{quant-ph/9505006}
\rightline{June 4, 1995}
}
\baselineskip 8mm

\begin{center}
{\bf Second solution of Demkov-Ostrovsky superpotentials}

\bigskip

Haret C. Rosu$^{a}$\cite{byline}, Marco Reyes$^{b}$\cite{byline},
Kurt Bernardo Wolf $^{c}$\cite{byline},
Octavio Obreg\'on$^{a,d}$\cite{byline}

{$^{a}$ \it Instituto de F\'{\i}sica de la Universidad de Guanajuato,
Apdo Postal E-143, L\'eon, Guanajuato, M\'exico}

{$^{b}$\it Departamento de F\'{\i}sica, Centro de Investigaci\'on y Estudios
Avanzados del Instituto Polit\'ecnico Nacional, Apdo Postal 14-740,
M\'exico Distrito Federal, M\'exico}

{$^{c}$\it Instituto de Investigaciones en Matem\'aticas Aplicadas y en
Sistemas-Cuernavaca, Universidad Nacional Aut\'onoma de M\'exico,
Apdo Postal 139-B, 62191 Cuernavaca, Morelos, M\'exico}

{$^{d}$\it
Universidad Aut\'onoma Metropolitana, Iztapalapa, Apdo Postal 55-534, 09340,
Distrito Federal, M\'exico}

\end{center}

\bigskip
\bigskip

\begin{abstract}
We work out the second solution of the Demkov-Ostrovsky superpotentials in
the $R_0=0$ sector.

\end{abstract}
\bigskip


In previous works \cite{do} we have presented a supersymmetric approach to
the Demkov-Ostrovsky (DO) focusing potentials at zero energy.
Following a remark of L\'evai \cite{lev}, we worked in the so-called $R_0$
sector defining the radial ground state and allowing to obtain the usual
Riccati equation for the superpotential
of the effective DO potentials in the Witten scheme.
The particular solution of the initial DO Riccati equation (DORE),
$W^2-dW/d\rho = U_{eff}^{DO}$,
was found by us to be $W_1=\frac{l}{\rho} -
\frac{(2l+1)}{\rho(1+\rho ^{2\kappa})}$, where
$\rho$ is a dimensionless radial coordinate, $l$ is the centrifugal quantum
number, and $\kappa$ is the DO index, with $\kappa$ =1 corresponding to the
Maxwell fisheye (MF) lens case and $\kappa$ =1/2 being the atomic Aufbau (AA)
case.

We want now to obtain the second solution of the initial DORE.
Suppose it to be $W={\cal V} ^{-1} + W_1$ \cite{n}, then by substituting in
the DORE one gets
$$\frac{d{\cal V}}{d\rho}+2W_1 {\cal V}=-1~.
\eqno(1) $$
This equation can be written as follows
$$ \frac{d}{d\rho}\Bigg[
{\cal V}\exp\left( 2\int W_1 d\rho\right)\Bigg]
=-\exp \left(2\int W_1 d\rho\right)~,
\eqno(2)$$
with the solution
$${\cal V} =-\exp\left(
-2\int W_1 d\rho\right)\cdot \int \exp\left(2\int W_1 d\rho\right) d\rho ~.
\eqno(3)$$
Since we know \cite{do} that $W_1=-\frac{d}{d\rho}\ln f(\rho)$ we
get $\int W_1 d\rho =-\ln f(\rho)$. Thus
$${\cal V}=-f^2(\rho)\int f^{-2}(\rho)d\rho ~.
\eqno(4)$$
The object of interest is now the integral of the inverse square of $f$, which
reads explicitly
$$\int \frac{(\rho ^{-\kappa}+\rho ^{\kappa})^{\frac{(2l+1)}{\kappa}}}
{\rho}d\rho~ .
\eqno(5)$$
Let $\rho ^{\kappa}=\tan (\frac{\alpha}{2})$. Then the integral turns into the
form
$$\frac{2^{\frac{2l+1}{\kappa}}}{\kappa}
\int \frac{d\alpha}{(\sin \alpha)^{(2l+\kappa +1)/\kappa}}~,
\eqno(6)$$
and for the cases of physical interest, MF and AA, the formulas
2.515.1 and 2.515.2, respectively, in Gradshteyn and Ryzhik \cite{gr}
should be used to express it as a series.

Thus, in the MF case, ($\kappa$ =1), the integral Eq.~(5)
can be worked out into the series
$$S_1=-\frac{2^{2l+1}}{2l+1}\cos \alpha \left \{(\csc \alpha)^{2l+1}+
\sum _{m=1}^l \frac{2^m[l(l-1)...(l+1-m)]}{[(2l-1)(2l-3)...(2l+1-2m)]}
(\csc \alpha)^{2l+1-2m}\right\}~,
\eqno(7)$$
while for the AA case, ($\kappa$ =1/2), the integral Eq.~(5) reads
$$S_{\frac{1}{2}}=
-\frac{2^{4l+3}}{2l+1}\cos \alpha(\csc ^{2} \alpha)^{(2l+1)}\left\{1 +
\sum_{m=1}^{2l}\frac{[(4l+1)(4l-1)...(4l-2m+3)]}
{(2\csc^{2} \alpha)^{m} [(2l)(2l-1)...(2l-m+1)]}
\right \}
+\frac{4[(4l+1)!!]}{4^{-l}[(2l+1)!]}\ln \tan (\frac{\alpha}{2})~,
\eqno(8)$$
where $\alpha =2\arctan \rho$ in the first formula and $\alpha=2\arctan
\sqrt{\rho}$ in the latter one.

In the MF case, the radial factor $f^2(\rho)$ can be written
trigonometrically as
$2^{-2l}(\sin\alpha)^{2l}\sin ^2(\frac{\alpha}{2})$ implying
$${\cal V} _{1}=\frac{2\cos \alpha}{2l+1}\tan (\frac{\alpha}{2})
\left\{1+
\sum _{m=1}^l \frac{(2\sin^2 \alpha)^m
[l(l-1)...(l-m)]}{[(2l-1)(2l-3)...(2l+1-2m)]}\right\}~.
\eqno(9)$$

In the AA case, the square radial factor is
$2^{-4l}(\sin \alpha)^{4l}\sin ^{4} (\frac{\alpha}{2})$ and
one can work out easily a formula for ${\cal V} _{\frac{1}{2}}$
$$
{\cal V} _{\frac{1}{2}}=  \frac{2\cos \alpha}{2l+1}
\tan ^2 (\frac{\alpha}{2})\left\{1+
\sum_{m=1}^{2l}\frac{(\frac{\sin^2 \alpha}{2})^m
[(4l+1)(4l-1)...(4l-2m+3)]}{(2l)(2l-1)...(2l-m+1)}
\right\}+ \frac{4[(4l+1)!!][\frac{\sin ^{2}(\alpha)}{2}]^{2l}}
{(2l+1)![\csc (\frac{\alpha}{2})]^{4}}
\ln \tan (\frac{\alpha}{2})~.
\eqno(10)$$
We have in this way all the ingredients for the second solution of DORE.
A more general solution contains a constant $\lambda$-parameter in Eq.~(3),
see \cite{n}, \cite{suk}. This allows a more direct connection with the
Gel'fand-Levitan inverse method \cite{n}.
The general solution means Eq.~(4) modified as follows
$$ {\cal V} _{\lambda,\kappa}=-f^2_{\kappa}(\rho)
\left(\lambda +\int _{\rho}^{\infty}f^{-2}_{\kappa}
(\rho ^{'})d\rho ^{'}\right)~.
\eqno(11) $$
In this case, the integral Eq.~(6) will have the lower limit $\alpha$ and
the upper
one $\pi$. Thus, ${\cal V} _{\lambda, \kappa}$ can be calculated from our
formulas as follows
${\cal V} _{\lambda, \kappa}={\cal V} _{\kappa}-\lambda f^{2}_{\kappa}(\rho)$.

\section*{Acknowledgments}
This work was partially supported by the CONACyT Projects 4862-E9406,
4868-E9406, and by the Project DGAPA IN 1042 93 at the Universidad
Nacional Autonoma de M\'exico.

M.R. was supported by a CONACyT Graduate Fellowship.



\end{document}